\documentclass[journal,twoside,hidelinks,a4paper]{IEEEtran}

\newcommand{\subparagraph}{}

\usepackage{amsmath}
\usepackage{rotating}
\usepackage{emptypage,titling,titlesec,widetext,textcomp,relsize,parskip}
\usepackage[utf8]{inputenc}
\usepackage[margin=10pt,font=footnotesize,labelsep=period]{caption}
\usepackage{tabulary,longtable,makecell,multirow,color,colortbl}
\usepackage[colorlinks=false,linkcolor=black,urlcolor=gray,citecolor=black]{hyperref}
\usepackage[usenames,dvipsnames,svgnames,table]{xcolor}
\usepackage{authblk,lipsum,txfonts,float,morefloats,gensymb,graphicx,setspace,fancyhdr,bibunits,booktabs,array,marvosym,xhfill}
\usepackage[normalem]{ulem}
\usepackage[export]{adjustbox}

\usepackage[includeheadfoot,paperwidth=8.5in,paperheight=11in,top=1cm,bottom=2cm,left=2cm,right=2cm,footskip=.25in,headheight=40pt,headsep=12pt,heightrounded]{geometry}

\newcommand{\etal}{\textit{et}~\textit{al.}~}
\newcommand{\ie}{\textit{i.e.,}~}
\newcommand{\eg}{\textit{e.g.,}~}
\newcommand{\resp}{\textit{resp.}~}
\newcommand{\etc}{\textit{etc.}}
\newcommand{\cf}{\textit{cf.}~}

\newcommand{\squeezeup}{\vspace{-3.5mm}}
\newcommand{\xfill}[2][1ex]{{\dimen0=#2\advance\dimen0 by #1\leaders\hrule height \dimen0 depth -#1\hfill}}
\newcommand{\xfilll}[2][1ex]{\dimen0=#2\advance\dimen0 by #1\leaders\hrule height \dimen0 depth -#1\hfill}

\renewcommand{\figurename}{Figure}

\titlespacing\section{0pt}{8pt plus 4pt minus 2pt}{2pt plus 2pt minus 2pt}
\titlespacing\subsection{0pt}{8pt plus 4pt minus 2pt}{2pt plus 2pt minus 2pt}
\titlespacing\subsubsection{0pt}{8pt plus 4pt minus 2pt}{2pt plus 2pt minus 2pt}

\newcommand{\subtitle}[1]{\posttitle{\par\end{center}\begin{center}\bfseries\Large #1\end{center}\vskip1.2em}}
\newcommand\myabstract[2][.8]{\renewcommand\maketitlehookd{\mbox{}\medskip\par\centering\begin{minipage}{#1\textwidth}\small#2\end{minipage}}}

\definecolor{myred}{RGB}{189, 3, 34}
\date{}

\title{\vspace{-1cm}{\includegraphics[height=0.5in,keepaspectratio=true]{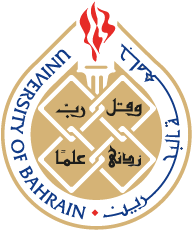}\hfill\parbox{15cm}{\raggedleft\sublargesize{\color{myred}\bfseries{International Journal of Computing and Digital Systems\\}}\sublargesize{ISSN (2210-142X)\\Int. J. Com. Dig. Sys. 14, No.1 (Jul-23)\\\vspace{3mm} \url{http://dx.doi.org/10.12785/ijcds/140123}}}\\\vspace{1cm}{\bfseries{\fontsize{18}{21.6}Requirements Traceability:}}}}

\subtitle{Recovering and Visualizing Traceability Links Between Requirements and Source Code of Object-oriented Software Systems}

\author[1]{\bfseries Ra'Fat Al-Msie'deen}

\affil[1]{\normalfont\textit{Department of Software Engineering, Faculty of IT, Mutah University, Mutah 61710, Karak, Jordan}}
\affil[ ]{\vspace{0pt}}
\affil[ ]{Received x Mon. 2023, Revised x Mon. 2023, Accepted x Mon. 2023, Published x Mon. 2023}
\affil[ ]{\vspace{0pt}}

\fancypagestyle{firstpage}{%
    \fancyhf{}
    \fancyhead[RO]{{\normalfont{Int. J. Com. Dig. Sys. 14, No.1, 279-295 (Jul-23)\hspace{4mm}}}\includegraphics[height=0.5in, keepaspectratio=true]{ub_logo}\hspace{3mm}\bfseries{\thepage}}
    \fancyhead[LE]{\bfseries{\thepage}\hspace{3mm}\includegraphics[height=0.5in,keepaspectratio=true]{ub_logo}\hspace{3mm}\normalfont\textit{Ra'Fat Al-Msie'deen: Recovering and Visualizing Traceability Links Between Requirements and Code...}}
    \fancyfoot[R]{\vspace{5mm}\textsl{\color{gray}{\textsf{\url{http://journals.uob.edu.bh}}}}}}

\fancypagestyle{plain}{%
    \fancyhf{}
    \fancyhead[LR]{\vspace{2.5cm}}
    \fancyfoot[L]{\vspace{0mm}{\small\textit{E-mail address: rafatalmsiedeen@mutah.edu.jo}}}
    \fancyfoot[R]{\vspace{5mm}\textsl{\color{gray}{\textsf{\url{http://journals.uob.edu.bh}}}}}}

\begin{document}
\bstctlcite{IEEEexample:BSTcontrol}
\myabstract[1]{\vspace{-1cm}\hrule\vspace{2mm}\textbf{Abstract:} Requirements traceability is an important activity to reach an effective requirements management method in the requirements engineering. \textit{Requirement-to-Code Traceability Links} (RtC-TLs) shape the relations between requirement and source code artifacts. RtC-TLs can assist engineers to know which parts of software code implement a specific requirement. In addition, these links can assist engineers to keep a correct mental model of software, and decreasing the risk of code quality degradation when requirements change with time mainly in large sized and complex software. However, manually recovering and preserving of these TLs puts an additional burden on engineers and is error-prone, tedious, and costly task. This paper introduces \textit{YamenTrace}, an automatic approach and implementation to recover and visualize RtC-TLs in Object-Oriented software based on \textit{Latent Semantic Indexing} (LSI) and \textit{Formal Concept Analysis} (FCA). The originality of YamenTrace is that it exploits all code identifier names, comments, and relations in TLs recovery process. YamenTrace uses LSI to find textual similarity across software code and requirements. While FCA employs to cluster similar code and requirements together. Furthermore, YamenTrace gives a visualization of recovered TLs. To validate YamenTrace, it applied on three case studies. The findings of this evaluation prove the importance and performance of YamenTrace proposal as most of RtC-TLs were correctly recovered and visualized.
\newline
\newline
\textbf{Keywords:} Software engineering, Requirements traceability, Requirements engineering, Formal concept analysis, Latent semantic indexing, Object-oriented source code. \vspace{1mm} \hrule}

\maketitle

\section{INTRODUCTION}
\label{sec:introduction_and_overview}

\textit{Requirements Engineering} (RE) aims at discovering, documenting, and maintaining a collection of requirements for the software system \cite{GrossmanF0478} \cite{DBZhao21}. RE involves five steps which are requirement discover, analysis, specification, validation, and management \cite{SommervilleIan}. \textit{Requirements Management} (RM) helps in maintaining requirement evolution during software development. RM is interested in all processes that lead to changing functional requirement of the software system \cite{RaFatRequirement}. \textit{Requirements Traceability} (RT) is the key activity of RM process. RM process aims at finding and maintaining a traceability link of a particular requirement from its origins (or sources), across its specification and development to its, consequent deployment and use, and over a cycles of continuous improvement and repetition in any of these stages \cite{TorkarRichard}.

In RE, RT is an important task to attain a successful and effective RM process \cite{DWhiteK22}. RtC-TLs shape the relations between requirement and source code artifacts. RtC-TLs can assist software engineers to know which segments of code implement a particular requirement. Recovering TLs between software \textit{Requirements and Source code} (RaS) is very useful in numerous \textit{Software Engineering} (SE) tasks such as software maintenance, reuse, and change \cite{Hanspeter} \cite{DBZhangTGH21} \cite{DBLbauer0MAGLE23}. Manually recovering and maintaining these TLs puts a further burden on engineers and is error-prone, tedious, costly mission, and some TLs may be missing. Traceability is a unique way to guarantee that the software code is consistent with its functional requirements. Traceability also ensures that software engineers have implemented all and only the required functional requirements \cite{NasirAliGA13}. This paper introduces \textit{YamenTrace}, an automatic approach to recover and visualize RtC-TLs in \textit{Object-Oriented} (OO) software system \cite{AlkkMsiedeenSHUV201516}.

The manual creation of TLs among software RaS is time consuming, error-prone, tedious task, and complex activity in the SE domain \cite{Spanoudakis}. Therefore, this study suggests mainly an automatic approach to recover TLs between RaS. TLs between RaS is an important process for software engineers to know which segments of code implement a certain requirement \cite{DRahimiC18}. Thus, when requirement changes are suggested software developers know which parts of software code have to be modified \cite{MoserAPSS0447}. Throughout software maintenance, a modification (or change) can not only influence source code but also cause an influence upon other artifacts such as requirements. Consequently, impact analysis \cite{WinAungHS20} can use TLs to comprehend relations and dependencies between software RaS.

\begin{figure}[H]
  \center
  \includegraphics[width=1\columnwidth]{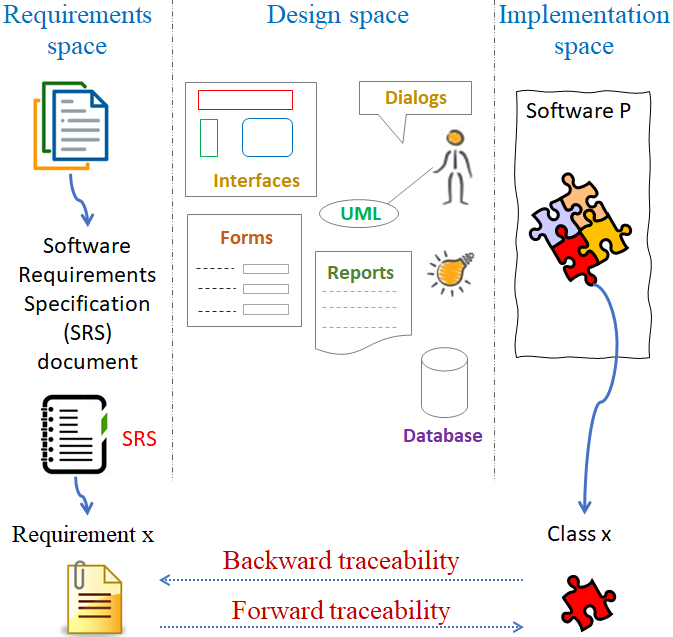}
  \caption{Backward and forward directions of requirement traces.}
  \label{Fig:TLsTypes}
\end{figure}

The traceability concept is defined as the degree of which every component in a software determines its reason for existing \cite{GotelCHZEGDAMM12}. Furthermore, traceability can defined as the degree of which a link can be formed among two or more software artifact \cite{IEEE15Glossary}. In this work, TLs is established between requirement and class documents of software system based on the \textit{Textual Similarity} (TS) between those documents. In RE domain, RT term is defined as the capacity to illustrate and follow the life cycle of a requirement, in both a forwards and backwards orientation \cite{GotelF94}. Therefore, the ability of software engineers to follow the traces to and from a requirement (\eg from its origins to its implementation) called RT \cite{WinklerP10}. \textit{Backward traceability} means the capacity to follow a TL from a particular software artifact to its origins (or sources) of which is has been developed (\eg code $\longrightarrow$ requirement). While \textit{forward traceability} \cite{IEEE278253} means the capacity to follow a TL from software artifact sources to its developed artifact (\eg requirement $\longrightarrow$ code). Figure \ref{Fig:TLsTypes} shows the Backward and forward directions of requirement traces. Software requirements define what the software should do. Functional requirements of software are statements of the services (or tasks) that the software must provide to its users \cite{ALmsiedeen12202111} \cite{RaFatbookModel}. In this work, functional requirements are described by natural language \cite{FRIJATAhmad}. Thus, for each requirement, there is a document describing single software service via short paragraph(s).

\textit{YamenTrace} combines two techniques in order to recover TLs between RaS. The first technique is LSI and the second one is FCA \cite{AlMsiedeenSHUVS13} \cite{ALMsiedeenGDRGPL}. \textit{Information Retrieval} (IR) techniques aim at identifying the documents that are relevant to a query in a group of documents \cite{LinLC22}. LSI is an IR technique. It uses \textit{Singular Value Decomposition} (SVD) on the \textit{Term-document Matrix} (TDM). In the scope of this study, LSI can be described as an IR technique that uses a set of documents as the inputs and produces an indicator with document similarities (\ie TS) as the output \cite{YurekliKB21}. FCA is a clustering technique. It allows to obtain an ordered collection of concepts from a \textit{dataset} consisted of objects expressed by attributes \cite{CarbonnelBHN20}. The researcher who is concerned with FCA and LSI can find additional information in several studies \cite{ALmsiedeen12} \cite{ALmsiedeen154} \cite{AlMsiedeenHSUV14}.

\begin{figure}[H]
  \center
  \includegraphics[width=.17\textwidth]{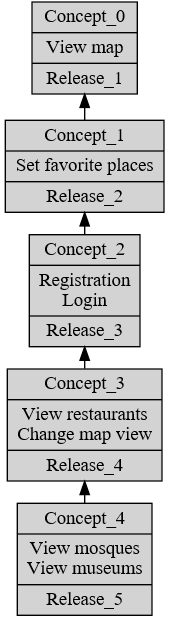}
  \caption{AOC-poset for FC of Table \ref{tab:JordanianMA}.}
  \label{Fig:FCAExaJMA}
\end{figure}

FCA considers as an important clustering technique in SE filed \cite{DBLPHaceneHNV10}. FCA enables software engineers to extract an ordered set of concepts (\ie C =\{C0, C1, C2, C3, C4, ...\}) from a considerable dataset. This dataset is called a \textit{Formal Context} (FC) which contained \textit{Objects} (O) expressed by \textit{Attributes} (A). An FC is a triple X = (O, A, BR) where BR is a \textit{Binary Relation} between O set and A set (\ie $BR \subseteq O × A$). An illustrative example of FC is presented in Table \ref{tab:JordanianMA}. This FC describing \textit{Jordan maps application} releases (\ie O) by their requirements (\ie A). A \textit{Formal Concept} (FO) is a pair (X, Y) made of an object set $X \subseteq O$ and their common attribute set $Y \subseteq A$. For the given concept C1 = (X1, Y1), X1 is the extent of the concept C1, while Y1 is the intent of the concept C1. In this paper, author will use the \textit{AOC-poset} for concepts visualization \cite{AlMsSUV14}. Figure \ref{Fig:FCAExaJMA} shows AOC-posets for FC of Table \ref{tab:JordanianMA}.

\begin{table}[H]
 \center
 \caption{FC describing Jordan maps application releases by their requirements.}
    \begin{tabular}{|c|c|c|c|c|c|c|c|c|}
    \toprule
\begin{sideways}              \end{sideways} &
\begin{sideways} Registration \end{sideways} &
\begin{sideways} Login\end{sideways} &
\begin{sideways} View map\end{sideways} &
\begin{sideways} View mosques\end{sideways} &
\begin{sideways} View restaurants\end{sideways} &
\begin{sideways} View museums\end{sideways} &
\begin{sideways} Change map view\end{sideways} &
\begin{sideways} Set favorite places\end{sideways}\\\hline

\textit{Release\_1}	&	&	&x	&	&	&	 &	 &   \\\hline
\textit{Release\_2}	&	&	&x	&	&	&	 &	 &  x\\\hline
\textit{Release\_3}	&x  &x	&x	&	&	&	 &	 &  x\\\hline
\textit{Release\_4}	&x  &x	&x  &	&x  &	 &	x&	x\\\hline
\textit{Release\_5}	&x	&x	&x	&x	&x  &	x&	x&	x\\
    \bottomrule
    \end{tabular}
 \label{tab:JordanianMA}
\end{table}

As concepts (\ie FOs) of AOC-posets are well ordered, the intent of the top concept  (\ie Concept\_0) contains requirements that are shared with all Jordan maps releases. While the intents of all remaining FOs (\ie Concept\_1 to Concept\_4) contain sets of requirements shared to a subset of Jordan maps releases but not all releases. Moreover, the extent of each concept in AOC-posets is the set of Jordan maps releases that shared these requirements. For example, the extent of Concept\_0 is "Release\_1", and the intent of this concept is "View map".

\begin{figure}[H]
  \center
  \includegraphics[width=1\columnwidth]{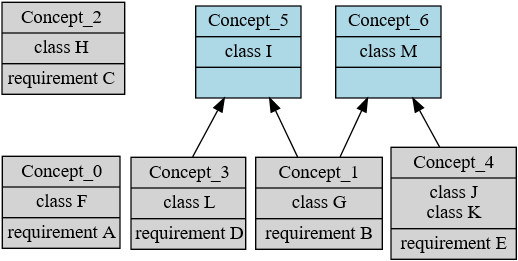}
  \caption{AOC-poset for FC of Table \ref{tab:requirementsImpl}.}
  \label{Fig:FCA_EXaRequirementsImpl}
\end{figure}

In this paper, after measuring TS between software requirement and code documents using LSI, \textit{YamenTrace} relies on FCA as a clustering technique to group similar documents together (\cf Figure \ref{Fig:FCA_EXaRequirementsImpl}). As another example regarding FCA technique, Table \ref{tab:requirementsImpl} shows a FC of a set of software requirements described by their implementation (\ie classes).

\begin{table}[H]
 \center
 \caption{FC describing software requirements by their implementation (\ie classes).}
    \begin{tabular}{|c|c|c|c|c|c|c|c|c|}
    \toprule
\begin{sideways}              \end{sideways} &
\begin{sideways}     class F  \end{sideways} &
\begin{sideways}     class G  \end{sideways} &	
\begin{sideways}     class H  \end{sideways} &
\begin{sideways}     class I  \end{sideways} &
\begin{sideways}     class J  \end{sideways} &
\begin{sideways}     class K  \end{sideways} &
\begin{sideways}     class L  \end{sideways} &
\begin{sideways}     class M \end{sideways}\\\hline
\textit{requirement A} &	x    &		  &        &	    &        &	      &        &         \\\hline		
\textit{requirement B} &	     & 	x     &        &        &        &        &        &  	x    \\\hline		
\textit{requirement C} &         &		  &	 	x  &  	x   &        &        &        &         \\\hline		
\textit{requirement D} &         &		  &		   &   	x   &        &        &    	x  &         \\\hline		
\textit{requirement E} &         &		  &		   &        &  	x    &   	x &        &   	x    \\		
    \bottomrule
    \end{tabular}
 \label{tab:requirementsImpl}
\end{table}

Figure \ref{Fig:FCA_EXaRequirementsImpl} displays AOC-posets for FC of Table \ref{tab:requirementsImpl}. Through this AOC-posets, we can notice that the Concept\_0 contains the requirement A in its extent, and class F in its intent. Thus, the requirement A is textually similar to the class F, and is therefore grouped together into a disjoint concept. Also, we can notice from this AOC-posets that the requirement B is textually similar to class G, class I, and class M. Moreover, class H is the implementation of requirement C. Similarly, requirement D implemented by class L and class I.

LSI refers to a technique that calculates TS between different documents. TS is calculated using the occurrences of terms in documents of the corpus. If several documents share a significant number of terms, then those documents are deemed to be similar \cite{jourigirCan}. A complete example explaining how to calculate TS between a set of documents using LSI technique is presented in Section \ref{subsec:LSI}.

\textit{YamenTrace} takes the software requirements and code as its inputs (\cf Figure \ref{Fig:RTrecoProcess}). Then, \textit{YamenTrace} recovers and visualizes the identified TLs between RaS. The first step of \textit{YamenTrace} aims at extracting software source code (\cf Section \ref{subsec:code}). The second step generates all class documents of a given software code (\cf Section \ref{subsec:codeDocuments}). Then, in the third step, \textit{YamenTrace} relies on LSI method to define the similarity between requirement and class documents (\cf Section \ref{subsec:LSI}). Finally, in the fourth step, \textit{YamenTrace} uses the similarity measure to identify TLs between RaS by using FCA (\cf Section \ref{subsec:TLsFCA}).

When software engineers maintain and evolve software system, RT becomes outdated because engineers don't care about updating traceability information. However, recovering RT later is a tedious and costly task for engineers. Thus, current studies have proposed several approaches to recover RT either semi-automatically or automatically (\cf Section \ref{sec:LITERATURE}).
Among the suggested studies, the current approaches revealed that IR methods can automatically retrieve TLs between RaS. Though, IR methods lack accuracy. This paper suggests an automatic approach to recovering and visualizing TLs between RaS based on LSI and FCA techniques.
The originality of \textit{YamenTrace} is that it exploits all code identifier names (\eg method and attribute), comments, and relations (\eg inheritance) in TLs recovery process. \textit{YamenTrace} uses LSI to find textual similarity across software code and requirements. While FCA employs to cluster similar code and requirements together. Furthermore, \textit{YamenTrace} gives a visualization of recovered TLs.

The rest of this paper is organized as follows. Studies relevant to \textit{YamenTrace} contributions are included in Section \ref{sec:LITERATURE}. \textit{YamenTrace} is detailed in Section \ref{sec:YAMENTRACE}. Experiments are shown in Section \ref{sec:EXPERIMENTATION}. Finally, Section \ref{sec:CONCLUSION} wraps up this paper and makes suggestions for future work.

\section{A LITERATURE REVIEW OF RT RECOVERY: A MINI SYSTEMATIC SURVEY}
\label{sec:LITERATURE}

This section presents a systematic literature review related to YamenTrace contributions. The closest approaches to \textit{YamenTrace} contributions are selected and presented in Table \ref{tab:approaches1}. In this Table, the used code relations, element names, and IR techniques (\resp FCA) are highlighted.

\begin{table*}[h]
 \center
 \caption{Summary of RT approaches (comparison table).}
    \begin{tabular}{|l|l|c|c|c|c|c|c|c|c|c|c|c|c|c|c|c|c|c|c|c|c|}
    \toprule
ID & Approach & \multicolumn{11}{c|}{Source code} &\multicolumn{6}{c|}{IR techniques} &  \\\cline{3-13}
   &	      & \multicolumn{3}{c|}{Code relations} & \multicolumn{8}{c|}{Element names} & \multicolumn{6}{c|}{} &
\begin{sideways} FCA \end{sideways} \\\cline{3-13} \cline{14-19}
&
&
\begin{sideways} Inheritance \end{sideways} &
\begin{sideways} Attribute access \end{sideways} &
\begin{sideways} Method invocation \end{sideways} &
\begin{sideways} Package \end{sideways} &
\begin{sideways} Class \end{sideways} &
\begin{sideways} Attribute \end{sideways} &
\begin{sideways} Method \end{sideways} &
\begin{sideways} Method parameter \end{sideways} &
\begin{sideways} Method local variable \end{sideways} &
\begin{sideways} Code Comment	  \end{sideways} &
\begin{sideways} Class file	  \end{sideways} &
\begin{sideways} LSI \end{sideways} &
\begin{sideways} VSM \end{sideways} &
\begin{sideways} PM \end{sideways} &
\begin{sideways} CML \end{sideways} &
\begin{sideways} JS \end{sideways} &
\begin{sideways} RTM \end{sideways}&
\\\hline
1&	Antoniol \etal \cite{AntoniolCCLM02} & & & & &x&x&x&x& &x& & &x&x& & &  &   \\\hline
2&	Tsuchiya \etal \cite{Tsuchiya}       & & & & &x& &x&x&x& & & & & &x& &  &	\\\hline
3&	Gethers \etal \cite{GethersOPL11}    & & & & & & & & & & &x& &x& & &x&x &   \\\hline
4&	Marcus \& Maletic \cite{MarcusM03}	 & & & & &x&x&x& & &x& &x& & & & & &	\\\hline			
5&	Ali \etal \cite{AliGA11Nasir}	     & & & & &x&x&x&x&x&x& & &x& & & & &	\\\hline			
$\rightsquigarrow$&	YamenTrace                           &x&x&x&x&x&x&x&x&x&x& &x& & & & & &x	\\	
    \bottomrule
    \end{tabular}
 \label{tab:approaches1}
\end{table*}

Antoniol \etal \cite{AntoniolCCLM02} presented an IR technique to retrieve TLs between software source code and documentation. They applied the suggested approach to trace software code to functional requirements (\resp manual pages). The authors used IR technologies, such as \textit{Probabilistic Model} (PM) and \textit{Vector Space Model} (VSM).

Tsuchiya \etal \cite{Tsuchiya} offered a semi-automatic approach to retrieve TLs between RaS in a collection of \textit{Product Variants} (PVs). The authors exploited commonality and variability at code elements and requirements to reduce the search space, then recover the TLs. The authors recovered TLs using the \textit{Configuration Management Log} (CML).

Gethers \etal \cite{GethersOPL11} presented an automatic approach to create a links between RaS of a single software system. Their approach combines several IR techniques such as: \textit{Jensen and Shannon} (JS) model, VSM and \textit{Relational Topic Modeling} (RTM). Their approach shows that the integrated technique outperforms separate IR techniques. The authors reached an average precision of about $40\%$. On the other hand, they did not offer specific values for recall metric.

Marcus and Maletic \cite{MarcusM03} used LSI technique for recovering TLs between software RaS. By using the identifier names and comments appear in software source code, they manage to mine semantic information valuable for retrieving TLs.

Ali \etal \cite{AliGA11Nasir} suggested an automatic IR approach in order to decrease the number of recovered false positive TLs by other IR studies. The suggested approach considers that information obtained from various code entities (\eg class and comments) are unique sources of information. Where, every source of information may serve as a specialist recommending TLs. The approach is used to decrease false positive TLs of VSM technique. The results reveal that the approach increases the accuracy of VSM, and it also decreases the efforts needed to manually eliminate false positive links. The current approaches for recovering TLs between RaS are summarized in Table \ref{tab:approaches1}.

Table \ref{tab:approaches2} shows a comparison between current approaches related to YamenTrace contributions. The selected approaches are evaluated according to the following criteria: link creation, tool support, empirical evidence, evaluation metrics, and code language.

\begin{table*}[h]
 \center
 \caption{An overview of RT approaches (comparison table).}
    \begin{tabular}{|c|l|c|c|c|c|c|c|c|c|c|c|c|}
    \toprule
 ID &	Approach & \multicolumn{2}{c|}{Link creation} &  & \multicolumn{3}{c|}{Empirical evidence}& \multicolumn{2}{c|}{Evaluation metrics} &\multicolumn{2}{c|}{Code language} \\ \cline{3-4}\cline{6-8} \cline{9-10} \cline{11-12}
  &
  &
\begin{sideways} Automatic\end{sideways} &	
\begin{sideways} Semi-automatic\end{sideways} &
\begin{sideways} Tool support\end{sideways} &
\begin{sideways} Academic\end{sideways} &	
\begin{sideways} Industrial\end{sideways} &	
\begin{sideways} Open Source\end{sideways} &
\begin{sideways} Precision\end{sideways} &	
\begin{sideways} Recall\end{sideways} &	
\begin{sideways}$C++$\end{sideways} &	
\begin{sideways}Java\end{sideways} \\\hline

1&	Antoniol \etal \cite{AntoniolCCLM02}&x& &x& &x& &x&x&x&x    \\\hline
2&	Tsuchiya \etal \cite{Tsuchiya}      & &x&x& &x& &x&x&x&x    \\\hline
3&	Gethers \etal \cite{GethersOPL11}   &x& &x&x& & &x& & &x    \\\hline
4&	Marcus and Maletic \cite{MarcusM03}	 &x& & & &x& &x&x&x&x    \\\hline			
5&	Ali \etal \cite{AliGA11Nasir}	     &x& &x& & &x&x&x&x&     \\\hline			
$--\rightsquigarrow$&	YamenTrace                           &x& & &x&x&x&x&x& &x    \\	

    \bottomrule
    \end{tabular}
 \label{tab:approaches2}
\end{table*}

Several studies have recovered TLs between requirements and a variety of software artifacts, such as design documents, \textit{Unified Modeling Language} (UML) diagrams.
Dagenais \etal \cite{DagenaisR12} have suggested a technique to extract TLs from API and learning sources based on code-like terms in documents. Their technique automatically analyzed the software documentation and linked code-like terms (\eg \textit{day()}) to explicit code elements (\eg \textit{DateTime.day()}) in the API.
Kaiya \etal \cite{KaiyaHaruhiko} have suggested a technique to discover change impacts on software code produced by requirements changes. They suggested a technique and a tool for impact analysis on source code produced by requirements changes. In their technique, an IR method is utilized to determine TLs between RaS.

Lin \etal \cite{LinLCSABBKDZ06} have introduced the \textit{Poirot} tool. This tool supports traceability of diverse software artifacts. A PM is utilized as an IR technique to automatically create traces among different kinds of software artifacts, involving software code and requirements. The authors did not employ precision (\resp recall) metric to evaluate the quality of the produced TLs.
Charrada \etal \cite{CharradaKG12} have suggested an approach to distinguishing outdated requirements by using changes in software source code. Their approach first finds changes in code that are likely to influence software requirements. Then it obtains a collection of keywords depicting changes. These keywords are tracked to \textit{Software Requirements Specification} (SRS) document to detect influenced requirements. The authors did not offer values for recall metric.

Yadla \etal \cite{YadlaHD05} have presented an approach to tracing software requirements to bug reports by using LSI. Their approach is implemented in the \textit{RETRO} tool. RETRO uses LSI technique to find TS among requirements and bug reports.
Eaddy \etal \cite{EaddyAAG08} have proposed \textit{CERBERUS}, a hybrid method for concern (or concept) location. The concern location problem is identifying code elements within a software that are relevant to the implementation of a feature or requirement. Their approach gave a value of 73\% (\resp 75\%) to the recall (\resp precision) metric.
Khetade and Nayyar \cite{KhetadePrashant} have suggested a method based on LSI to find TLs between software code and free text documents. They used LSI technique to automatically identify TLs among software code and requirements.

Eyal-Salman \etal \cite{SalmanSDA12} have suggested an approach based on LSI to recover TLs among source code and features of a collection of PVs. While YamenTrace recovers RtC-TLs in a single software system.
Chen \etal \cite{ChenHG12} \cite{ChenHGA18} have suggested an automatic approach that exploits hierarchical tree and tree map visualization techniques to offer a universal structure of requirement traces and a comprehensive overview of each trace. While YamenTrace provides graph-based visualization of RT information. This graph visualizes traceability information among software RaS.

A study of the literature and comparisons of current approaches showed that there is no study or approach in the literature that uses the code relations (\resp FCA) in the process of recovering TLs between software RaS.
In this paper, LSI is used to measure TS between requirement and class documents. The use of LSI in YamenTrace is not considered a novel aspect, as several studies have employed LSI to recover TLs between RaS.
On the other hand, FCA technique is used to cluster similar requirement and class documents together based on TS measured by LSI. The use of FCA here is considered a novel aspect of YamenTrace approach where it has not been used before in RT studies, especially in the context of RaS.
Also, YamenTrace prepares the class document in a novel way, where it exploits the identifier names, code relations and comments to construct the class document. Existing approaches used class files as it without any preprocessing.
Finally, YamenTrace visualizes the recovered TLs between RaS.

\section{YAMENTRACE APPROACH}
\label{sec:YAMENTRACE}

\begin{figure}[H]
  \center
  \includegraphics[width=0.95\columnwidth]{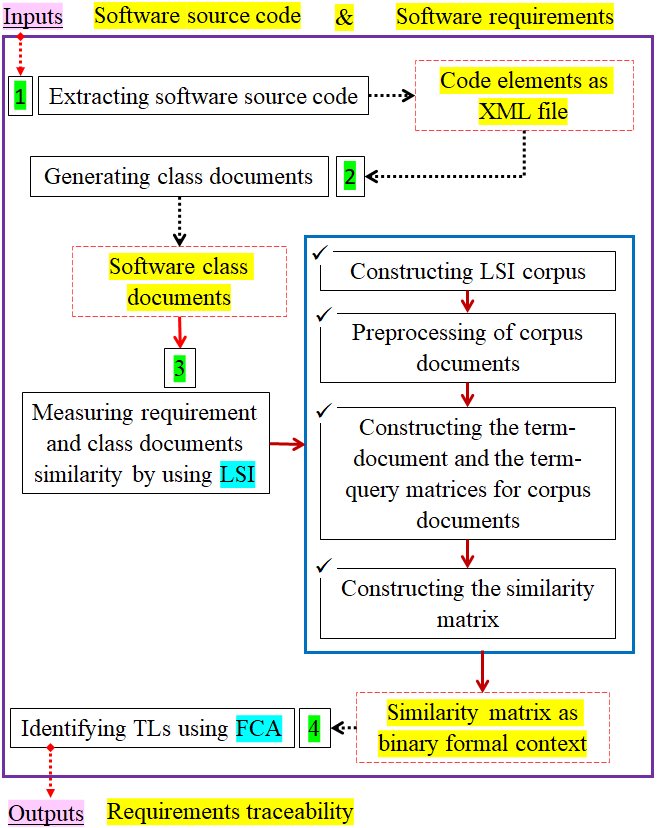}
  \caption{RT recovering process - \textit{YamenTrace} approach.}
  \label{Fig:RTrecoProcess}
\end{figure}

A summary of \textit{YamenTrace} is presented in Figure \ref{Fig:RTrecoProcess}. The inputs of \textit{YamenTrace} are software source code and requirements. While the outputs of \textit{YamenTrace} are TLs across software RaS.

This study focuses on the recovering of TLs between RaS. Functional requirements describe the software services that must present to end users. This study considers that functional requirements are implemented via software source code. \textit{YamenTrace} works only with OO software system \cite{FatMsiedeenDocumentation}. Consequently, functional requirements are implemented using main OO code elements (\eg class and method). This study also assumes that code identifiers that implement a particular requirement are textually similar to requirement name and description.

\begin{figure}[H]
  \center
  \includegraphics[width=0.85\columnwidth]{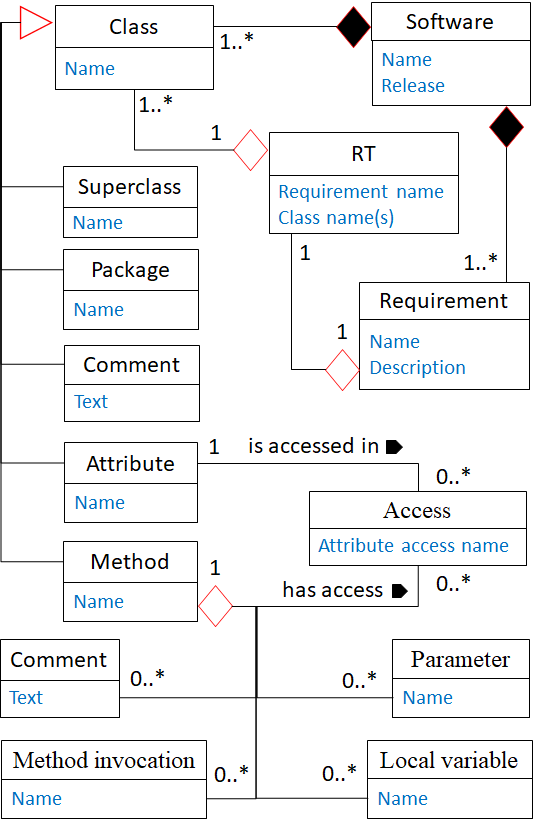}
  \caption{RT meta model of \textit{YamenTrace} approach.}
  \label{Fig:RTmetaModel}
\end{figure}

Figure \ref{Fig:RTmetaModel} displays a meta model representing the relation (or link) between software requirements and classes. In this work, requirement document is linked to one or more class documents based on the TS. Thus, for each requirement traceability, there is a requirement document and one or more class documents (\cf Figure \ref{Fig:RTmetaModel}). Each class document contains main source code elements that belong to this class. The software class may inherits attributes and methods from the superclass (\ie inheritance relation) \cite{DBLPAlMsiedeenSHUV13}. Also, each software class belongs to a particular software package. Also, software class contains many attributes, methods, and several code comments. In addition, software method may contains parameter name(s) in its signature. Also, in its body, the method may contains a local variable, code comment, attribute access, or method invocation. All code elements, code dependencies and requirement traceability relations are given in Figure \ref{Fig:RTmetaModel}.

\begin{figure}[H]
  \center
  \includegraphics[width=1\columnwidth]{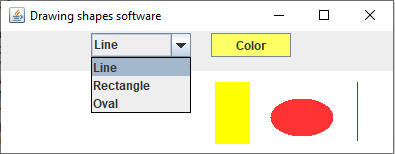}
  \caption{The GUI of DS application.}
  \label{Fig:DSa}
\end{figure}

To illustrate some steps of \textit{YamenTrace}, the author considers the \textit{Drawing Shapes} (DS) software as an illustrative example in this study \cite{ALmsiedeen125022}. DS permits the user to draw several types of shapes such as line, and rectangle. DS software is considered as a small sized software system \cite{drawingshapes}. The author uses this example to better clarify the steps of \textit{YamenTrace} approach. The author does not know the TLs between class and requirement documents in advance. \textit{YamenTrace} only uses software RaS as an inputs for RT recovering process. Figure \ref{Fig:DSa} displays the Graphical User Interface (GUI) of DS software.

DS allows user to choose a color and a kind of shape to be drawn from software interface. The possible shapes involve line, rectangle, and oval \cite{AuLabRafat} \cite{RaFatImpact}. The software engineer can extend this version by adding other kind of shapes. Furthermore, DS lets the user to press a mouse button to generate a shape on drawing zone. Users of this software can resize the drawn shape by dragging the mouse anywhere on the drawing zone. DS allows the user to draw a picture by mixing multiple shapes together. Table \ref{tab:requirementsOdSS} describes the functional requirements of DS software.

SRS document is the official document that contains functional (\resp non-functional) requirements of software system. This document contains the names of the requirements and a detailed description of each requirement. Natural language is usually used to describe a requirement of a program \cite{DBNayakTMPVP22}.

\begin{table}[H]
 \center
 \caption{Requirements and their description of DS software.}
    \begin{tabular}{|l|p{0.6\columnwidth}|}
    \toprule
\textbf{Requirement}    & \textbf{Requirement description} \\\hline
\textit{Draw a line}    & The software shall allow user to draw lines, and choose the right color of the drawn lines. Also, it shall allow end user to draw a single line or unlimited lines on the drawing zone. To draw a line, software shall provide a method like \textit{drawLine()} to draw a line between two points.\\\hline
\textit{Draw oval}      & The software shall allow user to draw ovals, and choose the right color of the drawn ovals. Also, it shall allow end user to draw a single oval or unlimited ovals on the drawing zone. To draw an oval, the software shall provide a method like \textit{drawOval()} to draw an oval.\\\hline
\textit{Draw rectangle} & The software shall allow user to draw rectangles, and choose the right color of the drawn rectangles. Also, it shall allow end user to draw a single rectangle or unlimited rectangles on the drawing zone. To draw a rectangle, the software shall provide a method like \textit{drawRectangle()} to draw a rectangle.\\
    \bottomrule
    \end{tabular}
 \label{tab:requirementsOdSS}
\end{table}

According to the suggested approach, \textit{YamenTrace} recovers TLs between software RaS in \textit{four} steps as described in the following.

\setlength{\parindent}{5mm}
 	
\subsection{Extracting software source code} \label{subsec:code}

\begin{figure}[H]
  \center
  \includegraphics[width=1\columnwidth]{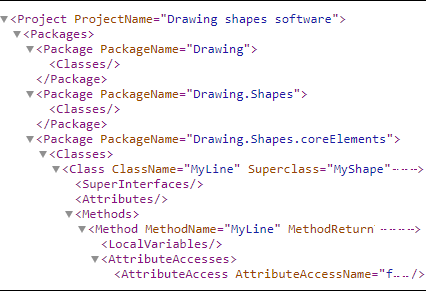}
  \caption{The code elements file of DS software (partial).}
  \label{Fig:xmlFile}
\end{figure}

The initial step of \textit{YamenTrace} is the extraction of software source code. Static code analysis \cite{AlMsieDeen1445} aims at identifying main OO elements (\eg class, method and comment). Static code analysis examines structural information (\eg data dependencies) of code \cite{RaFaBlasi}. For example, MyOval class extends MyShape class in DS software. This step takes software code as input and gives the code elements of software as output. \textit{YamenTrace} relies on this code elements to construct the class documents of the whole software. The main code elements such as class and method names are important sources of information in order to identify TLs between RaS. Figure \ref{Fig:xmlFile} shows the extracted code elements from DS software as XML file \cite{YamenTraceApproach}.

\subsection{Generating class documents} \label{subsec:codeDocuments}

\begin{figure}[H]
  \center
  \includegraphics[width=1\columnwidth]{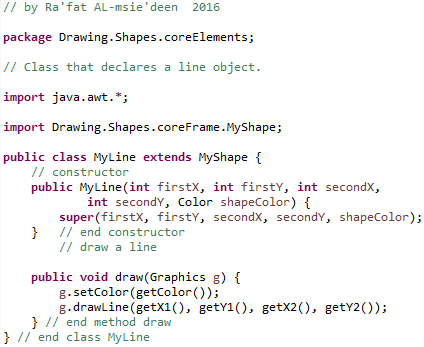}
  \caption{The code of \textit{MyLine} class from DS software.}
  \label{Fig:MyLineClass}
\end{figure}

In this step, \textit{YamenTrace} relies on the code elements file that is extracted in the previous step (\cf Section \ref{subsec:code}). \textit{YamenTrace} constructs the class documents for whole software based on code elements file. Each class document contains main code element names of this class, in addition to class and method comments (\resp. code relations). Figure \ref{Fig:MyLineClass} shows the code of \textit{MyLine} class from DS software.

\begin{figure}[H]
  \center
  \includegraphics[width=0.65\columnwidth]{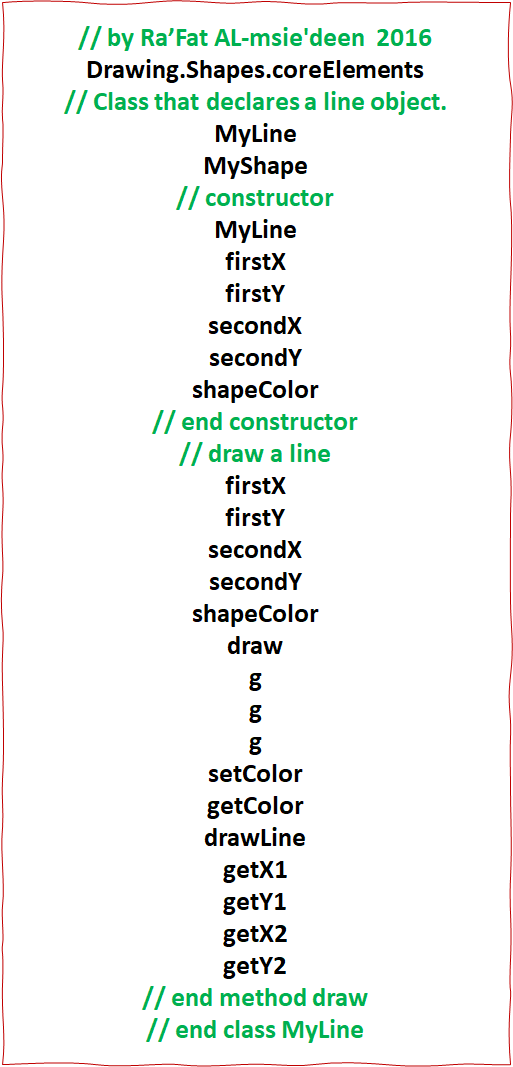}
  \caption{An example of a class document (\ie \textit{MyLine}) from DS software.}
  \label{Fig:ClassDocument}
\end{figure}

Figure \ref{Fig:ClassDocument} gives an example of the class document extracted from DS application. This document contains the package name (\ie \textit{Drawing.Shapes.coreElements}). This document also contains the class name (\ie \textit{MyLine}). In addition, it includes attribute and method names (\eg \textit{draw}). Also, from the method signature, it involves parameter names (\ie \textit{g}). Regarding the method body, it contains the local variable names. Also, class document contains code relation names such as: inheritance (\ie \textit{MyShape}), attribute access (\eg \textit{X1}) and method invocation (\eg \textit{drawLine}). Finally, the class document involves class and method comments (\eg \textit{$//$ draw a line}). Moreover, \textit{YamenTrace} names the document with the name of the class (\ie \textit{MyLine}).

\subsection{Measuring requirement and class documents similarity by using LSI} \label{subsec:LSI}

This paper considers that functional requirements are implemented by software classes. \textit{YamenTrace} bases the detection of subsets of software classes, which each implements a functional requirement, on the measurement of TS between these classes and software requirements. This similarity measure is determined based on LSI technique. \textit{YamenTrace} relies on the truth that classes engaged in implementing (or realizing) a functional requirement are textually nearer to one another than to the remainder of software classes. To calculate TS between software requirements and classes, \textit{YamenTrace} applied LSI technique in \textit{four} steps: 1) constructing LSI corpus, 2) preprocessing of corpus documents, 3) constructing TDM and Term-Query Matrix (TQM) and, finally, 4) constructing the \textit{Cosine Similarity Matrix} (CSM). The similarity of requirement and class documents is constructed with LSI as detailed in the following.

\subsubsection{Constructing LSI corpus}

LSI technique is a textual matching technique that aims to discover the TS between a query and a specified corpus of documents \cite{DBLPauerVE16}. A corpus represents a group of documents. In \textit{YamenTrace}, LSI corpus contains all software class documents. For query documents, each requirement document represents a query. The query document includes a description of a single software requirement, and it is named based on the name of that requirement. In \textit{YamenTrace}, the document-corpus contains all class documents, while the query-corpus contains all requirement documents. Table \ref{tab:DScorpus} offers the document and query corpus for DS software.

\begin{table}[H]
 \center
 \caption{Document and query corpus for DS software.}
    \begin{tabular}{|c|c|}
    \toprule
Query-corpus                &	Document-corpus \\
(\ie requirement documents) &	(\ie class documents) \\\hline
\textit{Draw a line}        &\textit{DrawingShapes}  \\\hline
\textit{Draw oval}          &\textit{MyLine}          \\\hline
\textit{Draw rectangle}     &\textit{MyOval}          \\\hline
                            &\textit{MyRectangle}     \\\cline{2-2}
                            &\textit{MyShape}          \\\cline{2-2}
                            &\textit{PaintJPanel}       \\
    \bottomrule
    \end{tabular}
 \label{tab:DScorpus}
\end{table}

\subsubsection{Preprocessing of corpus documents}

When corpus is created, the textual data of each document must be preprocessed in order to recover TLs between RaS. At the beginning, \textit{YamenTrace} removes stop words (\eg my, to, an, a, the, \etc), punctuation marks (\eg ?, !, \etc), special characters (\eg //, \&, \$, \etc), and numbers (\ie 0-9) from all corpus documents. Then, all document words are split into word tokens (\cf Table \ref{tab:myLineWT}) by using the camel-case syntax (\eg fillRect $\rightarrow$ fill and Rect). Camel-case is a generally applied technique for code splitting (or dividing) algorithms in the SE field \cite{RaFatAhmadDirasat}. Finally, \textit{YamenTrace} performs word stemming (\eg drawn $\rightarrow$ draw) on all document words (\cf Table \ref{tab:myLineWS}). In a SE domain, stemming (\eg removing word endings) is a text normalization method \cite{AlkkMsiedeenSHUV16} \cite{RaFat2019TagClouds}. In this step of YamenTrace, the word stemming is made by using \textit{WordNet} \cite{GeorgeMiller95}. All documents in document-corpus (\resp query-corpus) are preprocessed based on the previous procedure.

\begin{table}[H]
 \center
 \caption{Samples of the split word tokens from \textit{MyLine} class document.}
    \begin{tabular}{|c|c|}
    \toprule
Word                                 & Tokens                     \\\hline
\textit{shapeColor}                  & shape and color            \\\hline
\textit{getColor}                    & get and color              \\\hline
\textit{Shapes.coreElements}         & shapes, core, and elements \\
    \bottomrule
    \end{tabular}
 \label{tab:myLineWT}
\end{table}

This step aims at removing noise data from LSI corpus, saving memory space, and increasing the scale of \textit{YamenTrace} to work with the large sized software system. Thus, preprocessing helps software engineers to find better textual matching between RaS and improves the achieved results.

\begin{table}[H]
 \center
 \caption{Samples of the word stems (or roots) retrieved from \textit{MyLine} class document.}
    \begin{tabular}{|c|c|}
    \toprule
Word                       & Word stem (or root) \\\hline
\textit{Drawing}           & draw                \\\hline
\textit{Elements}          & element             \\\hline
\textit{Declares}          & declare             \\
    \bottomrule
    \end{tabular}
 \label{tab:myLineWS}
\end{table}

\subsubsection{Constructing TDM and TQM for corpus documents}

LSI technique starts with a TDM to count the occurrences of the \texttt{t} terms within a set of \texttt{d} documents. Thus, TDM is of the size \texttt{t × d} (\ie \texttt{TDM[t][d]}) where \texttt{t} (\resp \texttt{d}) is the number of unique-terms (\resp class documents) obtained from processed document-corpus (\cf Table \ref{tab:myLineTDMC}). In this matrix, each unique term (\resp class document) is denoted by a row (\resp column), with each matrix cell (\ie \texttt{TDM[t][d]}) representing an indicator of the weight of the \texttt{t}\emph{th} distinctive term in the \texttt{d}\emph{th} class document. The weight is really specified based on the value of term occurrence of the \texttt{t}\emph{th} term in the \texttt{d}\emph{th} class document \cite{AlMsieDeen1445}.

\begin{table}[H]
 \center
 \caption{TDM mined from document-corpus of DS software (partial).}
    \begin{tabular}{|l|c|c|c|c|c|c|}
    \toprule
 Term / Class                                &
\begin{sideways} DrawingShapes \end{sideways} &
\begin{sideways} MyLine \end{sideways} &
\begin{sideways} MyOval \end{sideways} &
\begin{sideways} MyRectangle \end{sideways} &
\begin{sideways} MyShape \end{sideways} &
\begin{sideways} PaintJPanel \end{sideways} \\\hline

\textit{line}	&0	&6	&0	&0	&0	&1  \\\hline
\textit{draw}	&1	&5	&3	&3	&2	&2  \\\hline
\textit{shape}	&21	&4	&3	&3	&6	&29 \\\hline
...	            &...&...&...&...&...&...\\
    \bottomrule
    \end{tabular}
 \label{tab:myLineTDMC}
\end{table}

In the same way of TDM, TQM is aimed to count the iterations of the \texttt{t} terms within a set of \texttt{q} query documents. Table \ref{tab:myLineTQMc} shows part of TQM mined from query-corpus of DS software.

\begin{table}[H]
 \center
 \caption{TQM mined from query-corpus of DS software (partial).}
    \begin{tabular}{|l|c|c|c|}
    \toprule
 Term / Requirement &
\begin{sideways} Draw a line \end{sideways} &
\begin{sideways} Draw oval \end{sideways} &
\begin{sideways} Draw rectangle \end{sideways} \\\hline

\textit{draw}	        &7	          &7	      &7  \\\hline
\textit{line}	        &7	          &0	      &0  \\\hline
\textit{shape}	        &0	          &0	      &0  \\\hline
...	                    &...	      &...        &... \\

    \bottomrule
    \end{tabular}
 \label{tab:myLineTQMc}
\end{table}

TQM is of the size \texttt{t × q} (\ie \texttt{TQM[t][q]}) where \texttt{t} (\resp \texttt{q}) is the number of unique-terms (\resp requirement documents) obtained from processed document-corpus (\resp query-corpus).

\subsubsection{Constructing the similarity matrix}

TS between query-corpus and document-corpus is represented by CSM. CSM columns and rows are represented as vectors of documents. The columns of CSM are documents of the document-corpus (\ie class documents), and the rows of CSM are documents of the query-corpus (\ie requirement documents). CSM cells take a value in a range between \texttt{-1} to \texttt{1} \cite{GrossmanF04}. Table \ref{tab:myLineSimilarityM} shows the extracted similarity matrix from requirement and class documents of DS software.

\begin{table*}[h]
 \center
 \caption{Similarity matrix for requirement and class documents of DS software.}
    \begin{tabular}{|l|c|c|c|c|c|c|}
    \toprule
	                         &DrawingShapes  &	  MyLine    &	MyOval      & MyRectangle   & MyShape       &PaintJPanel \\\hline
\textit{Draw a line}	     &0.037023712	 &0.989989848	&-0.131160318	&0.010703977	&-0.03271227	&0.012714135 \\\hline
\textit{Draw oval}	         &0.026077607	 &-0.070876593	&0.88846873	    &-0.452309648	&-0.011457777	&0.014299836 \\\hline
\textit{Draw rectangle}      &0.033470816	 &0.014789518	&-0.395774033	&0.9171953	    &-0.020937944	&0.018392209 \\
    \bottomrule
    \end{tabular}
 \label{tab:myLineSimilarityM}
\end{table*}

\subsection{Identifying TLs using FCA} \label{subsec:TLsFCA}

In this step, \textit{YamenTrace} uses FCA technique to recover, from requirement and class documents, which documents are textually similar. To convert the (numerical) CSM of the previous phase into (binary) FC, \textit{YamenTrace} utilizes a commonly used threshold for cosine similarity which is \texttt{0.70}. Thus, the only pairs of requirement and class documents having a counted similarity larger than or equal to the selected threshold (\ie \texttt{$\geq$0.70}) are deemed textually similar. Table \ref{tab:myLineSimilarityFC} illustrates the FC achieved by transforming CSM from Table \ref{tab:myLineSimilarityM} to binary FC.

\begin{table}[H]
 \center
 \caption{FC obtained from CSM in table \ref{tab:myLineSimilarityM}.}
    \begin{tabular}{|l|c|c|c|c|c|c|}
    \toprule
\begin{sideways}  \end{sideways} &
\begin{sideways} DrawingShapes \end{sideways} &
\begin{sideways} MyLine \end{sideways} &
\begin{sideways} MyOval  \end{sideways} &
\begin{sideways} MyRectangle \end{sideways} &
\begin{sideways} MyShape \end{sideways} &
\begin{sideways} PaintJPanel  \end{sideways}\\\hline

\textit{Draw a line}	    &0	  &1	&0	&0	 &0	      &0             \\\hline
\textit{Draw oval}	        &0	  &0	&1	&0	 &0	      &0             \\\hline
\textit{Draw rectangle}	    &0	  &0	&0	&1	 &0	      &0             \\
    \bottomrule
    \end{tabular}
 \label{tab:myLineSimilarityFC}
\end{table}

\begin{figure}[H]
  \center
  \includegraphics[width=1\columnwidth]{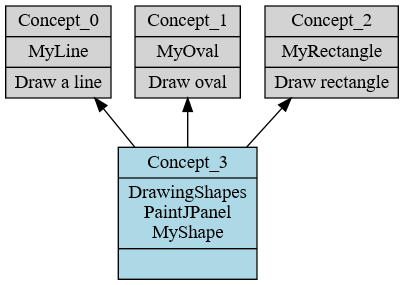}
  \caption{AOC-poset for FC of Table \ref{tab:myLineSimilarityFC}.}
  \label{Fig:AOC-posetDSs}
\end{figure}

As an instance, in FC of Table \ref{tab:myLineSimilarityFC}, the requirement query document "\emph{Draw a line}" is associated to the class document "\emph{MyLine}" since their similarity value equals \texttt{0.99}, which is larger than the chosen threshold (\ie \texttt{$\geq$0.70}). On the other hand, requirement query document "\emph{Draw a line}" and the class document "\emph{DrawingShapes}" are not associated since their TS equals \texttt{0.03}, which is fewer than the selected threshold (\ie \texttt{$\geq$0.70}). Figure \ref{Fig:AOC-posetDSs} shows the resulting AOC-poset from FC of Table \ref{tab:myLineSimilarityFC}.

AOC-poset in Figure \ref{Fig:AOC-posetDSs} displays \textit{four} concepts. Each concept of AOC-poset involves two elements: concept intent and extent \cite{Rafat2020}. Concept intent contains class documents, while concept extent contains requirement documents. For example, "\emph{draw a line}" requirement is textually similar to "\emph{MyLine}" class (\cf intent and extent of Concept\_0). Also, some class documents do not show any TS with any requirement documents (\cf intent of Concept\_3). Moreover, the intent of Concept\_2 (\ie \textit{MyRectangle}) is textually similar to the extent of the same concept (\ie \textit{Draw rectangle}).

\textit{YamenTrace} measures the quality and soundness of the recovered TLs using \textit{precision} and \textit{recall} measures. Precision and recall are two standard metrics widely employed in IR techniques \cite{Delater13}. Precision measure is the portion of recovered instances that are related (\cf Equation~\ref{eq1-Precision}), while recall measure is the portion of related instances that are recovered (\cf Equation~\ref{eq2-Recall}). The precision and recall metrics are computed as follows \cite{DBLPPorter22}:

\begin{equation} \label{eq1-Precision}
Precision  = \frac{RelatedLinks \cap RecoveredLinks}{RecoveredLinks}
\end{equation}

\begin{equation} \label{eq2-Recall}
Recall  = \frac{RelatedLinks \cap RecoveredLinks}{RelatedLinks}
\end{equation}

The most critical parameter to LSI technique is the number of selected \textit{Term-topics} (\texttt{T}). This parameter is called the \textit{Number of Topics} (\texttt{NoT}). In LSI, \texttt{T} is a set of terms that commonly co-occur in LSI corpus. \textit{YamenTrace} needs a sufficient \texttt{T} to obtain real term associations. \textit{YamenTrace} cannot employ a fixed \texttt{NoT} for LSI because it deals with several case studies of different sizes. For DS software case study, \texttt{NoT} is equal to \texttt{6}.

\begin{figure}[H]
  \center
  \includegraphics[width=0.90\columnwidth]{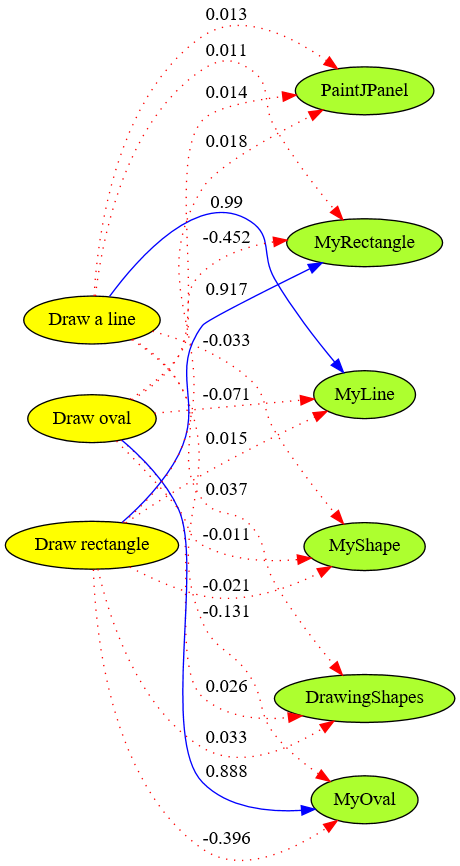}
  \caption{YamenTrace visualization of the recovered TLs from DS software.}
  \label{Fig:TLsVisualization}
\end{figure}

Figure \ref{Fig:TLsVisualization} shows the YamenTrace visualization of TLs between requirement and class documents for DS software. Visualizing TLs supports software engineers to recover and browse inter-relations among software documents in an intuitive manner \cite{DkbseMarcusXP05}.

\begin{figure}[H]
  \center
  \includegraphics[width=1\columnwidth]{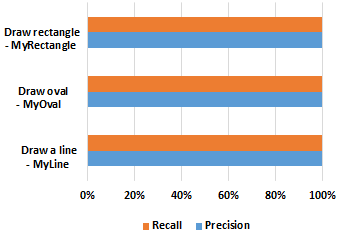}
  \caption{Evaluation metrics for the recovered TLs from DS software.}
  \label{Fig:PrecisionRecall}
\end{figure}

The author manually identified the correctly recovered TLs between software artifacts (\ie requirement and class documents) based on his excellent knowledge about DS software. Thus, RT information of DS software is well known and documented. RT information helps software developers or engineers in order to check and validate \textit{YamenTrace} findings. The success of \textit{YamenTrace} is determined by the values of the metrics of precision and recall. Each measure has a value between \texttt{0} and \texttt{1}. Figure \ref{Fig:PrecisionRecall} shows that recall measure is equal to \texttt{100\%} for all recovered TLs, which means that all related links are recovered. Also, Figure \ref{Fig:PrecisionRecall} shows that precision measure is equal to \texttt{100\%} for all recovered TLs, which means that all recovered links are related.

\textit{YamenTrace} obtained excellent results based on the evaluation metrics calculated for each \texttt{TL}. One explanation for this excellent finding is that a shared vocabulary is utilized in requirement descriptions and their implementations; therefore, TS was an appropriate way to identify TLs between RaS.

\section{EXPERIMENTATION}
\label{sec:EXPERIMENTATION}

This section presents the selected case studies, experimental results, evaluation metrics, implementation information, and threats to the validity of \textit{YamenTrace}.

Mobile Media (MM) is a Java open-source software system \cite{FigueiredoEduardo}. This software manipulates media (\eg photo) on mobile phones \cite{FigueiredoEduardo08}. The reason behind choosing MM as a case study is that this study is well documented and known. Also, software requirements and source code of MM are available freely online \cite{FigueiredoEduardo}. Moreover, the implementation of each requirement is well-known. Thus, MM (\ie release 8) artifacts are accessible for comparison with YamenTrace results, and to validate the approach proposal.

Health Watcher (HW) is a public health system \cite{HealthWatcher}. It is a Java open source software system \cite{HealthWatcher}. HW is a real health complaint software system \cite{Greenwood}. This software allows the citizen to register numerous types of health complaints (\eg complaints against food shops) \cite{DSoaresLB02}. Table \ref{tab:caseStudies} shows the standard software metrics for HW (\resp MM) software system. HW software considered as a large sized software system. HW software is well documented and known in SE filed. Requirements document of HW software is available for researchers \cite{RequirementsHealthWatcher}. In this case study, the implementation of each requirement is well known and documented.

DS, MM and HW software systems are presented in Table \ref{tab:caseStudies}. DS, MM and HW are described by the following metrics: Number of Software Packages (NOP), Number of Software Classes (NOC), Number of Software Attributes (NOA), Number of Software Methods (NOM), Number of Software Identifiers (NOI), Number of Software Comments (NOO), Number of  Local Variables (NOL), Number of Method Invocations (NOI), and Number of Attribute Accesses (NOE). All software metrics are extracted by the YamenTrace parser \cite{YamenTraceApproach}. The extracted XML code file for MM contains all needed code information for YamenTrace approach \cite{MMxmlFile}.

\begin{table}[H]
 \center
 \caption{DS, MM, and HW software metrics.}
 \scalebox{0.85}{
    \begin{tabular}{|l|c|c|c|c|c|c|c|c|c|c|}
    \toprule

\begin{sideways} Case study / Metric    \end{sideways} &
\begin{sideways} NOP                  \end{sideways} &
\begin{sideways} NOC                  \end{sideways} &
\begin{sideways} NOA                  \end{sideways} &
\begin{sideways} NOM                  \end{sideways} &
\begin{sideways} NOI                  \end{sideways} &	
\begin{sideways} NOO                  \end{sideways} &
\begin{sideways} NOL                  \end{sideways} &
\begin{sideways} NOI                  \end{sideways} &
\begin{sideways} NOE                  \end{sideways} \\\hline
\textit{DS}	              &4	&6	 &16  & 29	&55	 &112	&13	 & 99	&125   \\\hline
\textit{MM}	              &17	&51  &166 & 271	&505 &904	&258 &1200	&1790  \\\hline
\textit{HW}	              &22	&88	 &187 & 527	&824 &210	&524 &1952	&3303  \\
    \bottomrule
    \end{tabular}}
 \label{tab:caseStudies}
\end{table}

The number of requirements for MM (\resp HW) software is equal to 17 (\resp 9). Requirement names and descriptions are extracted from SRS document for each case study \cite{TizzeiLeonardo} \cite{RequirementsHealthWatcher}. Table \ref{tab:REaDesc} gives a samples of requirement names and their descriptions from MM and HW case studies. The complete set of requirements and their descriptions are available at YamenTrace webpage \cite{YamenTraceApproach}.

\begin{table*}[h]
 \center
 \caption{Samples of requirement names and their descriptions from MM and HW case studies.}
    \begin{tabular}{|c|l|p{0.66\textwidth}|}
    \toprule
\textbf{MM}&\textbf{Requirement}	          & \textbf{Requirement description}  \\\hline
Req01      &	\textit{Edit media label}     & Edit media label feature allows a user to label (or name) a photo (or media) with specific text. Labels or captions could be utilized for future search functionality.   \\\hline
Req02      &	\textit{Receive photo}	      & Receive photo via SMS message allows mobile user to receive a photo (or media) from other users by short messaging service. This requirement shall allow the SMS receiver controller to accept or reject the photo or media.   \\\hline
Req03      &	\textit{Exception handling}   & Exception handling is a non-functional requirement, and allows MM to handle exception related to media such as: image, photo album, and persistence exception.                   \\\hline
\textbf{HW}& Requirement	                  & Requirement description   \\\hline
Req01      &	\textit{Specify complaint}    &	This requirement lets a citizen to register complaints. Complaints can be: animal, food, or special complaint.  \\\hline
Req02      &	\textit{Register new employee}&	This requirement lets new employees to be registered on HW software.   \\\hline
Req03      &	\textit{Update employee}	  & This requirement lets of the employee's data or information to be updated on HW software. \\
    \bottomrule
    \end{tabular}
 \label{tab:REaDesc}
\end{table*}

\begin{figure}[H]
  \center
  \includegraphics[width=1\columnwidth]{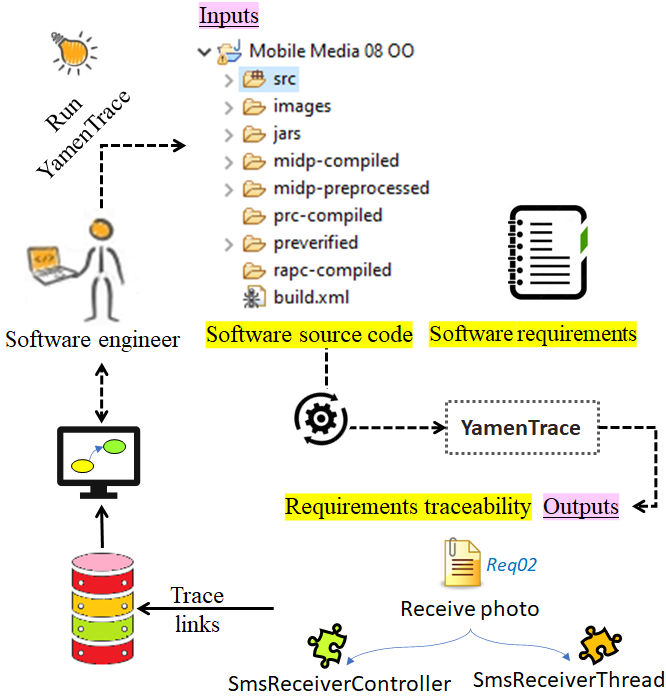}
  \caption{Running \textit{YamenTrace} approach on MM software.}
  \label{Fig:RunYamenTrace}
\end{figure}

Let's imagine that a software engineer is expected to trace a \textit{receive photo} requirement to its source code (\cf Figure \ref{Fig:RunYamenTrace}). The code is developed in Java language, and software requirements are created (or written) in English language (\cf Table \ref{tab:REaDesc}). Let's assume that software engineer is using YamenTrace, to identify TLs between requirements and code of MM case study. The software engineer is tracing a \textit{receive photo} requirement (\cf Req02 in Table \ref{tab:REaDesc}). Let's assume that the \textit{SmsReceiverThread} and \textit{SmsReceiverController} classes are the real implementation of \textit{receive photo}. In this case, engineer identifies a link via YamenTrace between Req02 to SmsReceiverThread and SmsReceiverController because the approach will find matched terms between Req02 and these two class documents. So, it is essential to consider all source code elements, comments, and relations in YamenTrace approach.

Table \ref{tab:RETLsResults} shows RtC-TLs results obtained from MM and HW software systems. Considering recall measure, its value is $100\%$ for all recovered TLs. This implies that all class documents that implement software requirements are recovered correctly. Also, results appear that precision metric seems to be high for some requirements and low for others. This means that not all recovered class documents are relevant to software requirements.

Results show that some class documents are associated with more than one requirement documents. For instance, \textit{addMediaToAlbum}, \textit{captureVideoScreen}, and \textit{videoCaptureController} class documents are linked to \textit{capture photo} and \textit{capture video} requirement documents. The reason behind this result is that capture photo and capture video requirements are implemented by same classes. Also, there are many textually matched terms between these documents.

\begin{table}[H]
 \center
 \caption{RtC-TLs results for MM and HW case studies.}
    \begin{tabular}{|c|l|c|c|}
    \toprule
\textbf{MM} &  \textbf{RtC-TL}  &\multicolumn{2}{c|}{\textbf{Evaluation Metrics}}\\\cline{3-4}
   &	                         & \textbf{Precision}& \textbf{Recall}   \\\hline
R01&	\textit{Create album}	             &55\%	     &100\%     \\\hline
R02&	\textit{Delete album}	             &80\%	     &100\%    \\\hline
R03&	\textit{Set favorite}	             &30\%	     &100\%    \\\hline
R04&	\textit{View favorite}	             &20\%	     &100\%    \\\hline
R05&	\textit{Sorting photo}	             &70\%	     &100\%    \\\hline
R06&	\textit{Create media}	             &60\%	     &100\%    \\\hline
R07&	\textit{Delete media}	             &80\%	     &100\%    \\\hline
R08&	\textit{Edit media label}	         &74\%	     &100\%    \\\hline
R09&	\textit{Copy media}	                 &50\%	     &100\%    \\\hline
R10&	\textit{Receive photo}	             &80\%	     &100\%    \\\hline
R11&	\textit{Send photo}	                 &80\%	     &100\%    \\\hline
R12&	\textit{View photo}	                 &40\%	     &100\%    \\\hline
R13&	\textit{Capture photo}	             &60\%	     &100\%    \\\hline
R14&	\textit{Play song}	                 &50\%	     &100\%    \\\hline
R15&	\textit{Play video}	                 &40\%	     &100\%    \\\hline
R16&	\textit{Capture video}	             &70\%	     &100\%    \\\hline
R17&	\textit{Exception handling}	         &60\%	     &100\%    \\\hline
\textbf{HW} & \textbf{RtC-TL}                &\textbf{Precision} &\textbf{Recall}  \\\hline
R01&	\textit{Query information}	         &50\%	     &100\%   \\\hline
R02&	\textit{Specify complaint}	         &80\%	     &100\%   \\\hline
R03&	\textit{Login}	                     &60\%	     &100\%   \\\hline
R04&	\textit{Register tables}             &65\%	     &100\%   \\\hline
R05&	\textit{Update complaint}	         &40\%	     &100\%   \\\hline
R06&	\textit{Register new employee}	     &30\%	     &100\%   \\\hline
R07&	\textit{Update employee}             &40\%	     &100\%   \\\hline
R08&	\textit{Update health unit}	         &50\%	     &100\%   \\\hline
R09&	\textit{Change logged employee}	     &70\%	     &100\%   \\
    \bottomrule
    \end{tabular}
 \label{tab:RETLsResults}
\end{table}

The results revealed that the use of all details from the class file, including identifier names, comments, and relations between code elements, caused a high value for the recall metric for all requirement documents. Thus, \textit{YamenTrace} approach is capable of identifying the real implementation of software requirements.
Furthermore, the results of the MM and HW software systems demonstrated that a single requirement can be implemented by one or more classes. On the other hand, one class can implement more than one requirement.

Moreover, results proved the ability of \textit{YamenTrace} to identify the implementation of functional and non-functional requirements. For instance, YamenTrace recovered the real implementation of the \textit{exception handling} requirement from MM case study.
Based on the manual analysis of the obtained RtC-TLs results, usually there is textual similarity between a requirement and its implementation. Generally, engineers name the code elements through which the requirement is implemented with a vocabulary similar to the requirement description (or name).
For each case study (\ie DS, MM, or HW), all experiment artifacts (\eg similarity matrix, AOC-poset, RtC-TLs visualization, \etc) are available on the YamenTrace webpage \cite{YamenTraceApproach}.

\noindent
\textbf{Implementation:} In order to recover TLs between RaS of a software system, \textit{YamenTrace} tool was developed and available on YamenTrace webpage \cite{YamenTraceApproach}. In order to extract the main OO elements, author has developed a code parser that depends on the \textit{Abstract Syntax Tree} (AST). AST is broadly employed in several areas of SE \cite{FischerFischerLG07}. AST is utilized as representation of software code. \textit{YamenTrace} parser uses the \textit{JDOM} library to extract the code elements in the form of an XML file. In order to apply LSI, the author has developed his LSI tool, which is available on YamenTrace webpage \cite{YamenTraceApproach}. For applying FCA, author used the \textit{Eclipse eRCA} \cite{eRCA}. Also, in order to visualize AOC-poset and recovered RtC-TLs, YamenTrace uses the \textit{Graphviz} library \cite{JohnEllsonGKNW04}.

\noindent
\textbf{Threats to validity:} \textit{YamenTrace} works only with Java applications. This considers a threat to implementation validity that restricts YamenTrace capability to work only with the applications that are written by Java language. Another threat to the validity of YamenTrace is that developer might not employ the same vocabularies used in requirement description to name source code elements that implement this requirement. This implies that TS may be not trustworthy (or should be enhanced with other methods) in all cases to recover TLs between software RaS.

\section{CONCLUSION AND PERSPECTIVES}
\label{sec:CONCLUSION}

\begin{figure}[H]
  \center
  \includegraphics[width=1\columnwidth]{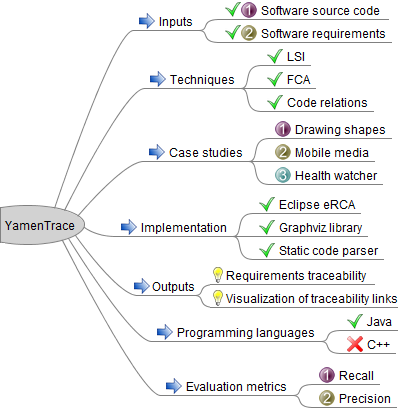}
  \caption{The key elements of \textit{YamenTrace} approach.}
  \label{Fig:basicElements}
\end{figure}

This paper suggested an approach based on LSI and FCA to recover and visualize RtC-TLs in a single software system. \textit{YamenTrace} has been implemented and evaluated on three case studies (\ie DS, MM, and HW). Findings displayed that most of RtC-TLs were recovered correctly. Figure \ref{Fig:basicElements} illustrates the key elements of YamenTrace approach.

The current approach works only with single software; therefor, one direction for future work of YamenTrace approach is to extend the current approach to work with a collection of PVs \cite{ALMsiedeenSHUVS13Seriai}. Then, it is important to extend the approach to identify the TLs between features and code of these PVs (\ie feature location) \cite{MarianneRafat2014}.

\textit{YamenTrace} can be extended in many ways. For instance, YamenTrace approach is designed for product written in Java language, thus, future work could aim on extending the current implementation of \textit{YamenTrace} to deal with other programming languages (\eg C++). Also, a further evaluation of \textit{YamenTrace} can be done with other case studies. To do this, it would be necessary to find suitable case studies whose requirements and source code are freely available to carry out the whole approach described in this paper.

\textit{YamenTrace} also plans to exploit useful information available in SRS document (\eg \textit{requirements dependency}) in TLs recovery process. Requirements dependency is an important aspect in tracing software requirements \cite{ASDeshpande22}. Furthermore, there is an urgent need to convert \textit{YamenTrace} to a generic approach, in order to be able to find TLs between any kind of software artifacts (\eg design documents, or features) and source code.

\bibliographystyle{IEEEtran}
\bibliography{references}
\vspace{-.5cm}
\begin{IEEEbiography}[{\includegraphics[width=1in,height=1.3in,clip,keepaspectratio]{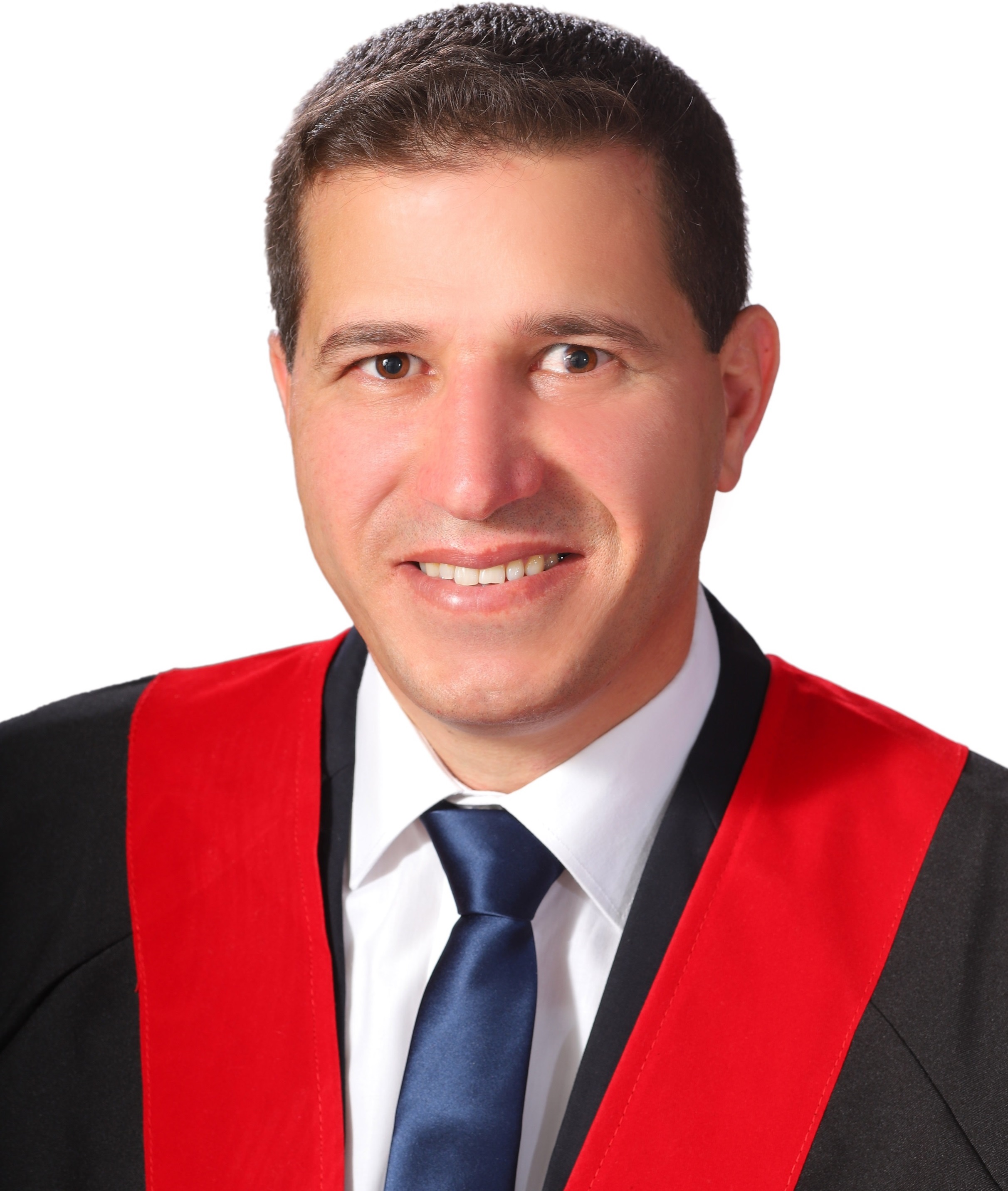}\squeezeup}]
{Ra'Fat Al-Msie'Deen} is an Associate Professor in the Software Engineering department at Mutah University since 2014. He received his PhD in Software Engineering from the Université de Montpellier, Montpellier - France, in 2014. He received his MSc in Information Technology from the University Utara Malaysia, Kedah - Malaysia, in 2009. He got his BSc in Computer Science from Al-Hussein Bin Talal University, Ma'an - Jordan, in 2007. His research interests include software engineering, requirements engineering, software product line engineering, feature identification, word clouds, and formal concept analysis.
\end{IEEEbiography}

\end{document}